\documentclass{PoS}
\usepackage{amsfonts}
\usepackage{mathrsfs}
\usepackage[ansinew]{inputenc}      
\usepackage{amssymb}                
\usepackage{amsmath}                
\usepackage[]{graphicx}
\usepackage{mathrsfs}
\usepackage{verbatim}
\usepackage{overpic}
\usepackage{xspace}
\usepackage{lineno}
\usepackage{enumitem}
\usepackage[font=small,skip=3mm]{caption}
\usepackage{setspace}
\usepackage{array}
\usepackage{lscape}
\usepackage{rotating}

\newcommand{\ttbar}{\ensuremath{t\bar{t}}\xspace}
\newcommand{\etmiss}{\ensuremath{E \kern-0.6em\slash_{\rm T}}\xspace}
\newcommand{\etmissx}{\ensuremath{E \kern-0.6em\slash_{\rm x}}\xspace}
\newcommand{\etmissy}{\ensuremath{E \kern-0.6em\slash_{\rm y}}\xspace}

\newcommand{\mlb}{\ensuremath{m_{\ell b}}\xspace}

\newcommand{\ljets}{\ensuremath{\ell\!+\!{\rm jets}}\xspace}
\newcommand{\dilep}{\ensuremath{\ell\ell}\xspace}

\newcommand{\etal}{\textit{et~al.}\xspace}

\newcommand{\GeV}{\ensuremath{\textnormal{GeV}}\xspace}
\newcommand{\TeV}{\ensuremath{\textnormal{TeV}}\xspace}

\newcommand{\dif}{\ensuremath{{\rm d}}}

\newcommand{\chisq}{\ensuremath{\chi^{2}}\xspace}

\newcommand{\met}{\ensuremath{E_\mathrm{T}^\mathrm{miss}}\xspace}

\newcommand{\fb}{\ensuremath{{\rm fb}^{-1}}\xspace}
\newcommand{\mw}{\ensuremath{M_W}\xspace}
\newcommand{\mt}{\ensuremath{m_t}\xspace}
\newcommand{\mtpole}{\ensuremath{m_t^{\rm pole}}\xspace}

\newcommand{\kjes}{\ensuremath{k_{\rm JES}}\xspace}

\newcommand{\pt}{\ensuremath{p_{\rm T}}\xspace}

\newcommand{\stt}{\ensuremath{\sigma_{t\bar t}}\xspace}
\newcommand{\stwo}{\ensuremath{\sqrt s=1.96~\TeV}\xspace}
\newcommand{\sseven}{\ensuremath{\sqrt s=7~\TeV}\xspace}
\newcommand{\seight}{\ensuremath{\sqrt s=8~\TeV}\xspace}

\newcommand{\sthirteen}{\ensuremath{\sqrt s=13~\TeV}\xspace}
\newcommand{\sttj}{\ensuremath{\sigma_{t\bar t+1~{\rm jet}}}\xspace}

\newcommand{\stat}{\ensuremath{{\rm(stat)}}\xspace}
\newcommand{\statjes}{\ensuremath{{\rm(stat\!+\!JES)}}\xspace}
\newcommand{\syst}{\ensuremath{{\rm(syst)}}\xspace}
\newcommand{\theo}{\ensuremath{{\rm(theo)}}\xspace}

\title{Measurements of the top quark mass from the LHC and the Tevatron}

\ShortTitle{Top quark mass measurements from LHC + Tevatron }

\author{\speaker{Oleg Brandt}\\
       {\bf{\sffamily on behalf of the ATLAS, CDF, CMS, and D0 Collaborations}}\\
        Universit\"at Heidelberg, Kirchhoff-Institut f\"ur Physik, INF 227,\\
        69120 Heidelberg, Germany\\
        E-mail: \email{oleg.brandt@kip.uni-heidelberg.de}}

\abstract{
The mass of the top quark is a fundamental parameter of the standard model and has to be determined experimentally. In these proceedings, I review recent measurements of the top quark mass in $pp$ collisions at $\sqrt s=7,~8,$ and 13~\TeV recorded by the ATLAS and CMS detectors at the LHC, and in $p\bar p$ collisions at $\sqrt s=1.96$~TeV recorded by the CDF and D0 experiments at the Tevatron. The measurements are performed in final states containing two, one, and no charged leptons. A relative precision of down to 0.3\% is attained. In addition, recent measurements aiming to determine the top quark mass in the well-defined pole scheme using both inclusive $\ttbar$ and $\ttbar+1~{\rm jet}$ production are presented.
}

\FullConference{ 9th International Workshop on the CKM Unitarity Triangle\\
         28  November - 3 December 2016\\
         Tata Institute for Fundamental Research (TIFR), Mumbai, India}

\begin{document}

\section{Introduction}
Since its discovery~\cite{bib:discoverydzero,bib:discoverycdf}, the determination of the top quark mass \mt, a fundamental parameter of the standard model (SM), has been one of the main goals of the CERN Large Hadron Collider (LHC) and of the Fermilab Tevatron Collider. Indeed, \mt and masses of $W$ and Higgs bosons are related through radiative corrections that provide a consistency check of the SM~\cite{bib:lepewwg,bib:theory}. Furthermore, \mt dominantly affects the stability of the SM Higgs potential~\cite{bib:theory,bib:vstab1}.
With $\mt=173.34\pm0.76~\GeV$, a world-average combined precision of 0.44\% has been achieved~\cite{bib:combiworld}.

In the SM, the top quark decays to a $W$~boson and a $b$~quark nearly 100\% of the time.
Thus, $\ttbar$ events are classified according to $W$ boson decays as ``dileptonic''~(\dilep), ``lepton+jets'' (\ljets), or ``all--jets''. Single top production contributes significantly at the LHC through the $qg\to q't\bar b$ process. In the following, I will present representative measurements in the three channels; a full listing of \mt results from the LHC and the Tevatron can be accessed through Refs.~\cite{bib:topresatlas,bib:toprescdf,bib:toprescms,bib:topresdzero}.

\section{Standard measurements of the top quark mass} \label{sec:standard}

\begin{figure}[b]
\centering
\begin{overpic}[clip,height=4.5cm]{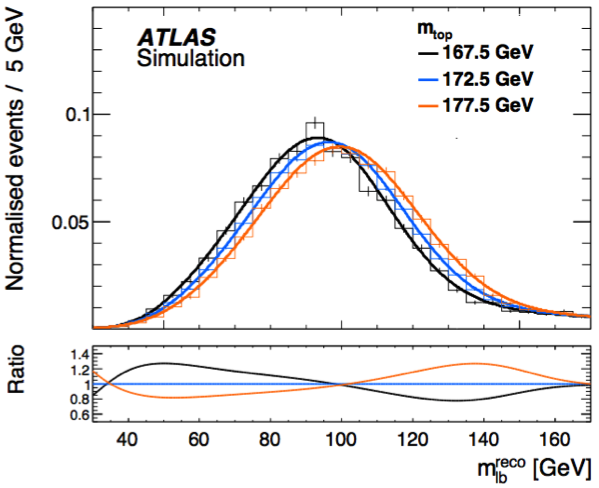}
\put(20,62){\bf\sffamily{(a)}}
\end{overpic}
\qquad
\begin{overpic}[clip,height=4.5cm]{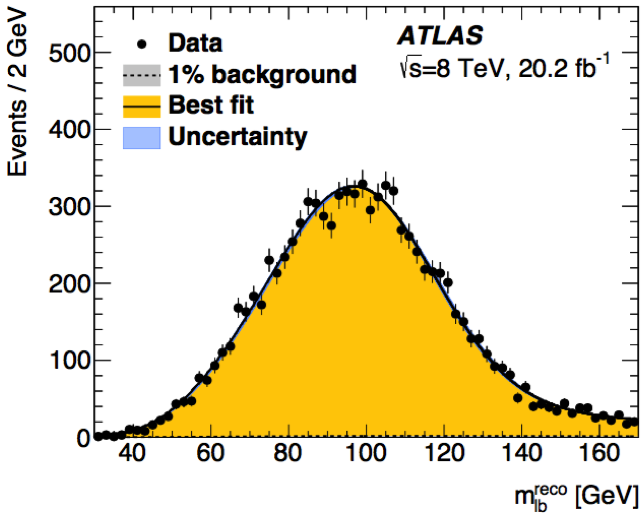}
\put(20,48){\bf\sffamily{(b)}}
\end{overpic}
\caption{
\label{fig:ll}
{\bf(a)} Expected dependence of the \mlb distribution of processes involving top quarks on \mt from Monte Carlo simulations at \seight with the ATLAS detector~\cite{bib:a8_ll}. 
{\bf(b)} The distribution in $\mlb$ in 20.3~\fb of data at \seight with the ATLAS detector. The predictions correspond to the best-fit values.
}
\end{figure}

The most precise single measurement of \mt in the \dilep channel is performed by the ATLAS Collaboration using 20.2~\fb of $pp$ collisions at \seight~\cite{bib:a8_ll}. The selection requires two isolated leptons ($e$ or $\mu$) of opposite charge, missing transverse momentum \met due to neutrinos, and $\geq 2$ jets, where at least one of which is identified as originating from a $b$ quark ($b$-tagged). A transverse momentum $p_{T,\ell b}>120~\GeV$ is required for the average of the two $\ell b$ systems to reduce the dominant uncertainty from the jet energy scale (JES). The \mt is extracted with the ``template method'', which in this case fits the distribution in the average invariant mass of the $\ell b$ system to the expectations from Monte Carlo (MC) simulations for different \mt, shown in Fig.~\ref{fig:ll}~(a). The best fit to data is shown in Fig.~\ref{fig:ll}~(b), and results in $\mt=172.99\pm0.41\stat\pm0.74\syst$~GeV. 
Tevatron's most precise single measurement in the \dilep channel of $m_t=173.32\pm1.36\stat\pm0.85\syst~\GeV$ is performed by the D0 Collaboration using 9.7~\fb of $p\bar p$ collisions at \stwo~\cite{bib:d0_ll}.

\begin{figure}
\centering
\begin{overpic}[clip,height=4.5cm]{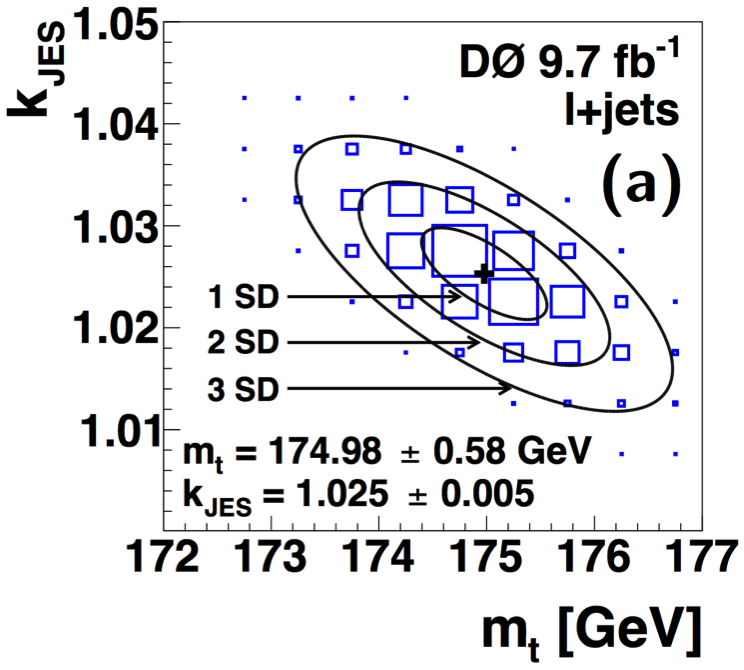}
\end{overpic}
\qquad
\begin{overpic}[clip,height=4.5cm]{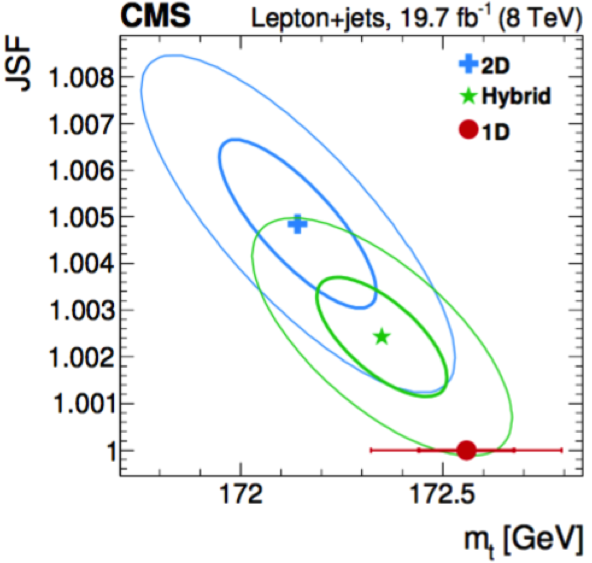}
\put(80,62){\large\bf\sffamily{(b)}}
\end{overpic}
\caption{
\label{fig:lj}
{\bf(a)} The likelihood in $(\mt,\kjes)$ in 9.7~\fb of $p\bar p$ collisions at \stwo recorded with the D0 detector~\cite{bib:d0_lj}. Fitted contours of equal probability are overlaid as solid lines. The maximum is marked with a cross. 
{\bf(b)}~Same as (a), but in 19.5~\fb of $pp$ collisions at \seight recorded with the CMS detector~\cite{bib:c8_lj}. The central result corresponds to ``Hybrid'', and \kjes is denoted as ``JSF''.
}
\end{figure}

The most precise single measurement of \mt from the Tevatron is performed by the D0 Collaboration using 9.7~\fb of data in the \ljets channel~\cite{bib:d0_lj} with a ``matrix element~(ME) method''. This approach determines the probability of observing a given event under both the $\ttbar$ signal and background hypotheses, as a function of \mt. This probability is calculated {\em ab initio} using the respective MEs of the \ttbar signal and dominant $W$+jets background, taking into account effects from parton showering (PS), hadronisation, and finite detector resolution.
This selection requires the presence of one isolated lepton, \met, and exactly four jets with at least one $b$-tag. A new JES calibration from exclusive $\gamma+$jet, $Z+$jet, and dijet events is applied to account for differences in detector response to jets originating from a gluon, a $b$~quark, and $u,d,s,$ or $c$~quarks. The overall JES \kjes is calibrated {\it in situ} by constraining the reconstructed invariant mass of the hadronically decaying $W$ boson to $\mw=80.4$~GeV. The likelihood over all candidate events is maximised in $(\mt,\kjes)$ as shown in Fig.~\ref{fig:lj}~(a), and $\mt=174.98\pm0.58\statjes\pm0.49\syst~\GeV$ is obtained. The most precise \mt result from the CDF Collaboration in the \ljets channel of $\mt=172.85\pm0.71\statjes\pm0.85\syst~\GeV$~\cite{bib:cdf_lj} is obtained with the template method.

The most precise single measurement of \mt from the LHC is performed by the CMS Collaboration using 19.7~\fb of data at \seight in the \ljets channel~\cite{bib:c8_lj}. The analysis uses a similar selection to the D0 result and applies the ``ideogramm method'' to extract \mt. Similar to the ME method, this approach calculates the probability to observe a given event as a function of $(\mt,\kjes)$. However, this probability is not calculated {\em ab initio}, but is obtained from MC simulations, in analogy to the template method. The final result of $\mt=172.35\pm0.16\statjes\pm0.48\syst$~GeV is represented in Fig.~\ref{fig:lj}~(b). The most precise \mt result from the ATLAS Collaboration is obtained with the template method using 4.7~\fb of data at \sseven and reads $\mt=172.33\pm0.75\statjes\pm1.02\syst~\GeV$~\cite{bib:a7_lj}.


The all-jets channel is particularly challenging due to very high background from QCD multijets. Tevatron's most precise single \mt result in this channel comes from the CDF Collaboration using 9.3~\fb of data~\cite{bib:cdf_jj}. A neural network and $b$-tagging enhance the signal-to-background ratio from $10^{-3}$ to about 1. The correct assignment of jets to partons is determined by minimising a $\chi^2$, which accounts for consistency of the two dijet systems with $m_W$, consistency of the two $jjb$ systems with each other, and consistency of the individual fitted jet momenta with measured ones, within experimental resolutions. The measured value is $\mt=175.07\pm1.19\statjes\pm1.55\syst~\GeV$. The most precise result in the all-jets channel at the LHC of $\mt=172.32\pm0.25\statjes\pm0.59\syst~\GeV$ comes from the CMS Collaboration~\cite{bib:c8_lj}.

\begin{figure}
\centering
\begin{overpic}[clip,height=10cm]{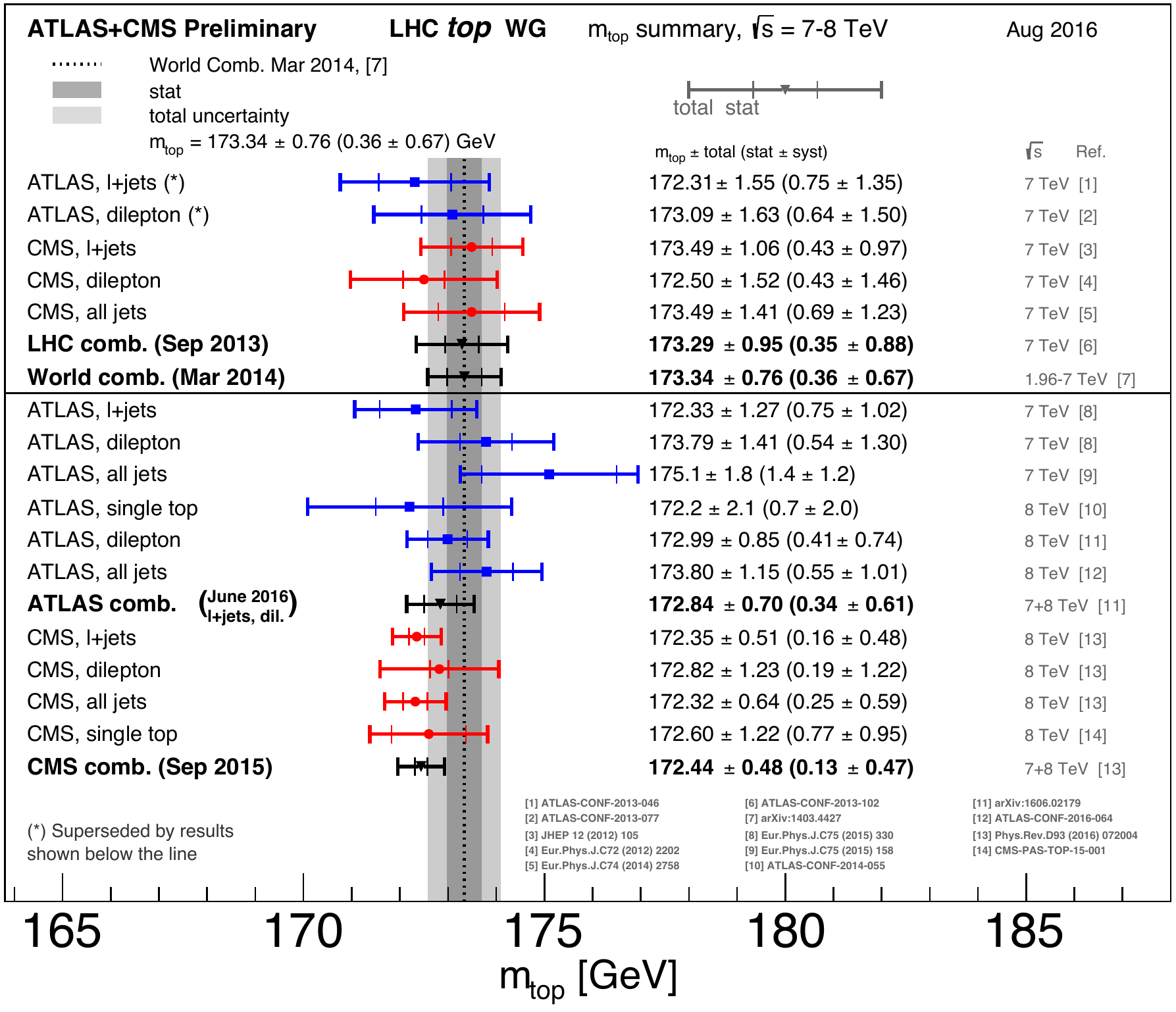}
\end{overpic}
\caption{
\label{fig:overview}
Overview of recent \mt measurements at the LHC~\cite{bib:overview_LHC}. References to the individual measurements are given at the bottom of the Figure.
}
\end{figure}

An overview of recent \mt measurements at the LHC~\cite{bib:overview_LHC} is given in Fig.~\ref{fig:overview}. A combination of \mt measurements from Run~I and II of the Tevatron considering statistical and systematic correlations yields $\mt=174.30\pm0.35\stat\pm0.34\syst~\GeV$~\cite{bib:combo_Tevatron}.

\section{Measurements of the top quark mass in the pole scheme} \label{sec:mtpole_tt}

The {\em standard} measurements of \mt from Sect.~\ref{sec:standard} are experimentally the most precise ones. However, they extract an \mt~{\em parameter} as implemented in MC generators, which is related to the pole mass scheme definition \mtpole in the SM Lagrangian within an uncertainty of $\leq$1~GeV~\cite{bib:alt}. 

The first LHC result on \mt at \sthirteen is an extraction of \mtpole from \stt performed by CMS in the \ljets channel using 2.3~\fb of data~\cite{bib:mtpole_tt}. This analysis exploits the dependence of \stt on \mtpole, which is now known with $\approx$3\% precision at NNLO with NNLL corrections~\cite{bib:xsec_tt_nnlo}. The input measurement of \stt achieves a relative uncertainty of $\approx$4\% by constraining the dominant $W$+jets background through sidebands in low jet and $b$-tag multiplicities, and using the difference in $\dif\sigma/\dif m_{\ell b}$ dependence between signal and background. The final result is $\mtpole=173.3^{+2.3}_{-2.0}({\rm stat+syst})\,^{+1.6}_{-1.1}\theo~\GeV$.

\begin{figure}
\centering
\begin{overpic}[clip,height=5.0cm]{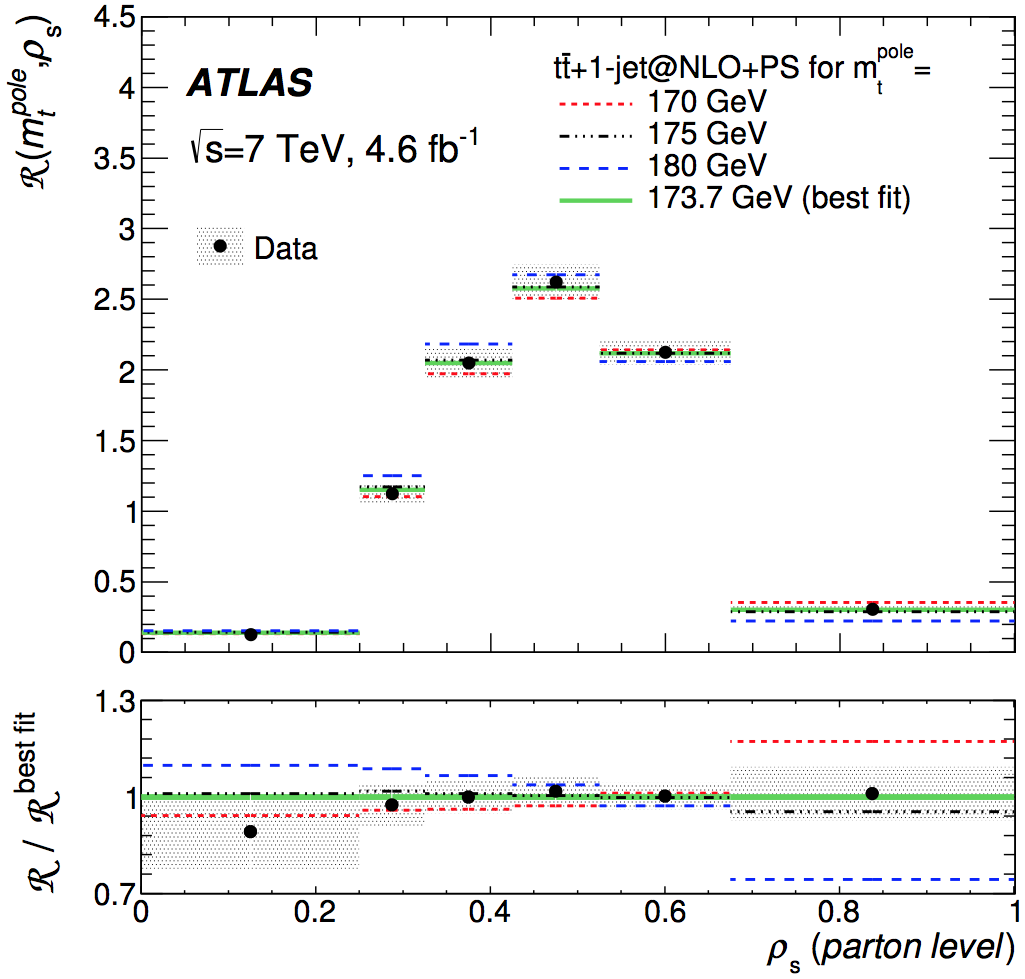}
\put(35,86){\bf\sffamily{(a)}}
\end{overpic}
\qquad
\begin{overpic}[clip,height=5.0cm]{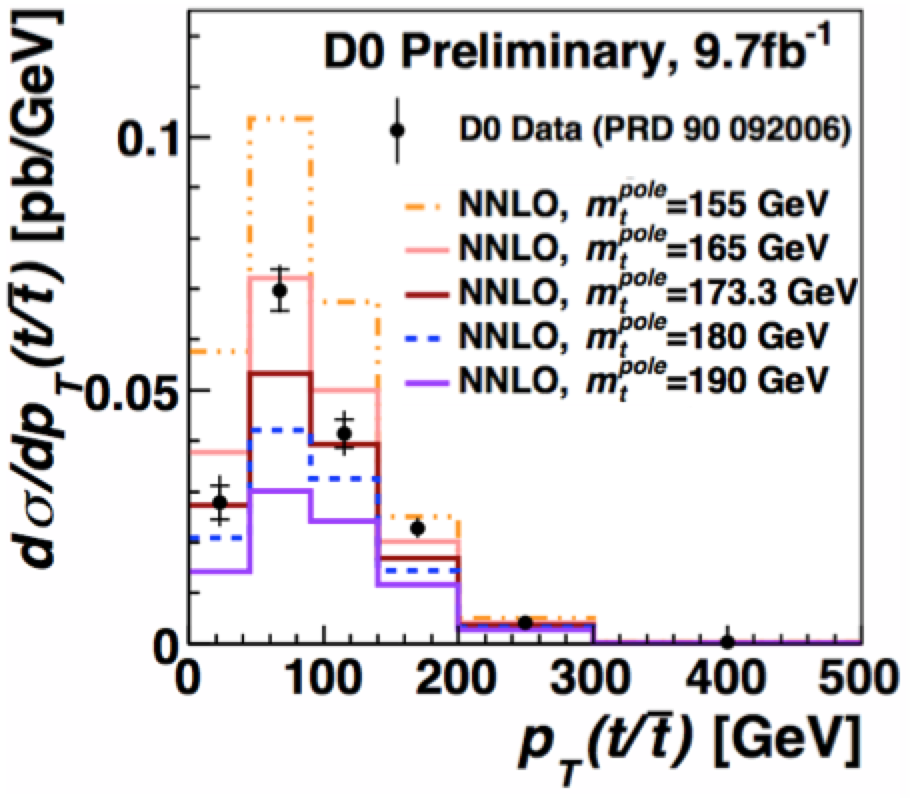}
\put(24,80){\bf\sffamily{(b)}}
\end{overpic}
\caption{
\label{fig:mtpole}
{\bf(a)}~The distribution $\mathcal R\equiv1/\sigma_{\ttbar+1\rm jet} \cdot \dif \sigma_{\ttbar+1\rm jet}/\dif \rho_s$ in $pp$ collisions at \sseven with the ATLAS detector~\cite{bib:mtpole_ttj}, compared to NLO predictions~\cite{bib:mtpole_ttj_xsec}. 
{\bf(b)} The distribution of $\dif\stt/\dif p_{T,t/\bar t}(\mt)$ in $p\bar p$ collisions at \stwo with the D0 detector~\cite{bib:mtpole_ttdiff}, compared to NNLO predictions~\cite{bib:mtpole_ttdiff_xsec}. 
Both distributions are shown  at parton level, after corrections for detector, PS, and hadronisation effects.
}
\end{figure}

The most precise \mtpole measurement is performed by the ATLAS Collaboration in the \ljets channel using 4.6~\fb of data at \sseven~\cite{bib:mtpole_ttj}. The \mtpole is extracted from from the production cross section of a \ttbar system in association with a jet \sttj, since the radiation rate of a high-\pt gluon off the \ttbar system is proportional to \mtpole. More precisely, the differential production cross section $\mathcal R(\mtpole,\rho_s)\equiv1/\sigma_{\ttbar+1\rm jet} \cdot \dif \sigma_{\ttbar+1\rm jet}/\dif \rho_s$ is compared to NLO calculations~\cite{bib:mtpole_ttj_xsec}, where $\rho_s \equiv 2m_0/\sqrt{s_{\ttbar+1{\rm jet}}}$, and the arbitrary constant $m_0$ is set to 170~GeV in this analysis. The selection is similar to other analyses in the \ljets channel discussed in Sect.~\ref{sec:standard}, and the correct jet-parton assignment is determined through a $\chi^2$ kinematic fit. To reduce the total uncertainty, $\pt>50~\GeV$ is required for the extra jet. The distribution in $\rho_s$ is corrected for detector, PS, hadronisation effects, and the presence of background. The resulting distribution at parton level is given in Fig.~\ref{fig:mtpole}~(a). The final result reads $\mtpole=173.1\pm1.50\stat\pm1.43\syst^{+0.93}_{-0.49}\theo~\GeV$. 

The second most precise \mtpole measurement is performed by the D0 Collaboration in the \ljets channel using 9.7~\fb of data~\cite{bib:mtpole_ttdiff}. This analysis extracts \mtpole by relating measured $\dif\stt/\dif m_{\ttbar}(\mt)$ and $\dif\stt/\dif p_{T,t/\bar t}(\mt)$ to recent NNLO and NLO calculations~\cite{bib:mtpole_ttdiff_xsec}. {\em Differential} cross sections allow for a more complete use of kinematic information, and thus a notably higher statistical precision than the \mtpole extraction from an inclusive \stt measurement. The selection is similar to Ref.~\cite{bib:d0_lj}, and the correct jet-parton assignment is identified through a \chisq kinematic fit. The resulting distributions are corrected for detector, PS, hadronisation effects, and the presence of background to obtain $\dif\stt/\dif m_{\ttbar}(\mt)$ and $\dif\stt/\dif p_{T,t/\bar t}(\mt)$, which are then directly compared to theory calculations to extract \mtpole. The final result reads $\mtpole=169.1\pm2.5({\rm stat+syst})\pm1.5\theo~\GeV$.

\section{Conclusions}
I presented recent measurements of the top quark mass, a fundamental parameter of the SM. The most precise single measurements at the LHC and the Tevatron of respectively $\mt=172.35\pm0.16\statjes\pm0.48\syst$~GeV and  $\mt=174.98\pm0.58\statjes\pm0.49\syst~\GeV$ are performed by the CMS and D0 Collaborations in the \ljets channel, corresponding to a relative precision of 0.30\% and 0.43\%. The precision of \mt measurements in the pole scheme is improved to 1.3\% due to the advent of new theory calculations and experimental approaches.

I would like to thank my colleagues from the ATLAS, CDF, CMS, and D0 experiments for their help in preparing this article, the staffs at CERN and Fermilab together with their collaborating institutions, as well as the relevant funding agencies.


\newpage
\begin{thebibliography}{99}

\bibitem{bib:discoverydzero}
S. Abachi \etal (D0 Coll.),
Phys. Rev. Lett. \textbf{74} (1995) 2632.

\bibitem{bib:discoverycdf}
F. Abe \etal (CDF Coll.), 
Phys. Rev. Lett. \textbf{74} (1995) 2626.

\bibitem{bib:lepewwg}
The ALEPH, CDF, D0, DELPHI, L3, OPAL, SLC Coll., 
\href{http://arxiv.org/abs/1012.2367}{arXiv:1012.2367 [hep-ex]} (2010).

\bibitem{bib:theory}
Rohini Godbole, these proceedings.

\bibitem{bib:vstab1}
G. Degrassi \etal, 
J. High Energy Phys. {\bf 08} (2012) 098;
F. Bezrukov \etal, 
Phys. Lett. B {\bf 659} (2008) 703;
A. De Simone \etal,
Phys. Lett. B {\bf 678} (2009) 1.

%


\bibitem{bib:combiworld}
ATLAS, CDF, CMS, and D0 Coll.,
\href{http://arxiv.org/abs/1403.4427}{arXiv:1403.4427 [hep-ex]} (2014).


\bibitem{bib:topresatlas}
\href{https://twiki.cern.ch/twiki/bin/view/AtlasPublic/TopPublicResults}{https://twiki.cern.ch/twiki/bin/view/AtlasPublic/TopPublicResults}.

\bibitem{bib:toprescdf}
\href{https://www-cdf.fnal.gov/physics/new/top/public_mass.html}{https://www-cdf.fnal.gov/physics/new/top/public\_mass.html}.

\bibitem{bib:toprescms}
\href{https://twiki.cern.ch/twiki/bin/view/CMSPublic/PhysicsResultsTOP}{https://twiki.cern.ch/twiki/bin/view/CMSPublic/PhysicsResultsTOP}.

\bibitem{bib:topresdzero}
\href{https://www-d0.fnal.gov/Run2Physics/top/top_public_web_pages/top_public.html#mass}{https://www-d0.fnal.gov/Run2Physics/top/top\_public\_web\_pages/top\_public.html\#mass}.




\bibitem{bib:a8_ll}
ATLAS Coll., Phys. Lett. B {\bf 761} (2016) 350.

\bibitem{bib:d0_ll}
D0 Coll.,
Phys. Lett. B {\bf 752} (2016) 18.

\bibitem{bib:d0_lj}
D0 Coll.,
Phys. Rev. Lett. {\bf 113} (2014) 032002.

\bibitem{bib:cdf_lj}
CDF Coll.,
Phys. Rev. Lett. {\bf 113} (2014) 032002.

\bibitem{bib:c8_lj}
CMS Coll.,
Phys. Rev. D {\bf 93} (2016) 072004.

\bibitem{bib:a7_lj}
ATLAS Coll., 
Eur. Phys. J. C {\bf 75} (2015), 330.

\bibitem{bib:cdf_jj}
CDF Coll.,
Phys. Rev. D {\bf 90} (2014) 091101.

\bibitem{bib:overview_LHC}
ATLAS and CMS Coll.,
\href{https://atlas.web.cern.ch/Atlas/GROUPS/PHYSICS/CombinedSummaryPlots/TOP/mtopSummary_TopLHC/history.html}{https://atlas.web.cern.ch/Atlas/GROUPS/PHYSICS/\\CombinedSummaryPlots/TOP/mtopSummary\_TopLHC/history.html}

\bibitem{bib:combo_Tevatron}
CDF and D0 Coll.,
\href{https://arxiv.org/abs/1608.01881}{arXiv:1608.01881 [hep-ex]}.

\bibitem{bib:alt}
A. H. Hoang and I. W. Stewart, 
Nucl. Phys. Proc. Suppl. {\bf 185} (2008) 220.

\bibitem{bib:mtpole_tt}
CMS Coll.,
\href{http://cms-results.web.cern.ch/cms-results/public-results/preliminary-results/TOP-16-006/index.html}{CMS-PAS-TOP-16-006} (2016).

\bibitem{bib:xsec_tt_nnlo}
M. Czakon \etal,
Phys. Rev. Lett. {\bf 110} (2013) 252004.

\bibitem{bib:mtpole_ttj}
ATLAS Coll., 
J. of High Energy Phys. {\bf 10} (2015) 121.

\bibitem{bib:mtpole_ttj_xsec}
S. Alioli \etal,
Eur. Phys. J. C {\bf 73} (2013) 2438.

\bibitem{bib:mtpole_ttdiff}
D0 Coll.,
\href{https://www-d0.fnal.gov/Run2Physics/WWW/results/prelim/TOP/T113/}{D0 CONF Note 6473} (2016).

\bibitem{bib:mtpole_ttdiff_xsec}
M. Czakon, P. Fiedler, D. Heymes and A. Mitov, 
J. of High Energy Phys. {\bf 05} (2016) 034.

\end{thebibliography}
\end{document}